**Possible Origin of Gravity and the Holographic Principle**

Arnold Stein[*]

Department of Biological Sciences, Purdue University, West Lafayette, IN 47907, USA

*E-mail address: steina@purdue.edu

**Abstract**

Assuming the holographic principle, the gravitational force can be formulated thermodynamically as an entropic force, but the mechanism by which the attraction between two masses occurs is not clear. The physical basis for the holographic principle is also unknown. My primary assumption is that empty space consists of discrete countable units that have on the order of one bit of entropy per cubic Planck unit. The basic idea here is simply that rather than matter just occupying empty space, the presence of matter excludes discrete units of space that have entropy. I argue that this volume exclusion of empty space leads to an attractive gravitational force and to the holographic principle.

Gravity, one of the four forces of nature, is responsible for the evolution of the universe, and is a pervasive force experienced by everyone and everything. The gravitational force was described empirically by Newton [1] more than three hundred years ago. It was then described geometrically, nearly one hundred years ago, by Einstein in terms of the curvature of space (or, more precisely, spacetime) [2]. Despite these advances, the gravitational force has proven difficult to incorporate into quantum mechanics, which appears to describe everything else extremely well. The application of Einstein's field equations to black holes led to the realization that the entropy of the volume of space closely surrounding a black hole, which was presumed to contain the highest entropy possible, is proportional to the surface area, rather than the volume as would be expected for a volume of gas molecules or for a box of marbles [3,4]. This curious result was further developed into the holographic principle, which essentially states that the number of bits of information within a volume of space cannot be greater than four times the area of the surrounding surface, measured in Planck units [5-7]. Interestingly, by assuming the holographic principle, it is possible to derive Einstein's field equations [8], and, very simply, Newton's gravitational law [9]. In this letter, I suggest that the discreteness of space can lead to the gravitational force and to the holographic principle. I sketch how these conclusions can be deduced.

It is clear that the gravitational force can be formulated thermodynamically as an entropic force [8-10], but the mechanism by which the attraction between two masses occurs is not clear. The tendency of the ends of a polymer chain to come together [9] is not really an appropriate model for how two unconnected masses attract each other in space. Also, how an imagined holographic screen in space can serve as a semi-permeable membrane [9] is not very clear. Some other entropic mechanism must be operative.

The holographic principle strongly suggests that empty space consists of discrete units on the Planck scale, Planck volumes [11]. The Planck length is about $10^{-33}$ cm, and a Planck volume is about $10^{-99}$ cm$^3$. Empty space has generally been regarded as containing no entropy. However, if empty space really consists of discrete countable units, it should have entropy. I will assume that this entropy is on the order of 1 bit/cubic Planck unit. This is my primary assumption. The shapes of the unit Planck volumes, their packing arrangement, packing density, quantum fluctuations, and the quantum physics of their interaction with matter remain to be determined. The basic idea here is simply that rather than matter just occupying empty space, the presence of matter in space excludes discrete units of space which have entropy. I will argue that volume exclusion of empty space then leads to an attractive gravitational force and to the holographic principle.

Although it is not very well known, volume exclusion of small particles by larger ones is predicted to generate an entropic attractive force between the larger particles. This result was predicted theoretically in 1958 by Asakura and Oosawa [12], and was verified experimentally in the 1990's [13]. There is now growing evidence that this mechanism is important in driving cellular organization in biological systems [14]. The idea is simple (see ref. 14, Figure 1 for an illustrative figure). Two large particles, existing within a volume containing many small particles that are in random motion, will tend to attract each other because the large particles exclude volume to the small particles, and when the large particles come together they exclude less volume than when they are apart (because their excluded volumes then overlap). Thus, there is more volume accessible to the (much greater number of) small particles, increasing the entropy of the system. An alternative equivalent way of looking at this phenomenon is that an osmotic pressure, due to the random motion of the small particles, drives the two large particles together because the osmotic pressure becomes progressively lower in the region between the large particles as they approach each other and exclude progressively more small particles selectively from the region between them. Thus, the anisotropic bombardment of the large particles (when they are near each other) by the small ones drives the large particles together.

This phenomenon is easy to demonstrate. Figure 1 (**a-h**) shows time points of a movie, at the times indicated. A 20 cm diameter Pyrex glass dish containing approximately 37,500 small (2 mm diameter) glass beads and two 2.5 cm diameter marbles, placed 10.5 cm apart, was shaken on a rotary shaker to modestly agitate the glass beads. After times ranging from about 20 seconds to a few minutes (Figure 1 **i**), the two marbles generally spiraled together as illustrated in Figure 1 **j**, and then remained together rotating about each other.

To show that the marbles did not come together because of the possible existence of a region in the dish that the marbles preferred to go to, the same two marbles were shaken individually in the same dish, under the same conditions and starting at the same locations as in the previous experiment. The distance between the two marble locations was measured every ten seconds over five minutes. The results from ten experiments were averaged and compared with the results of ten experiments for which the two marbles were shaken together. Figure 1 **k** shows that the individually shaken marbles did not tend to occupy the same dish locations at the same time to any appreciable extent, whereas the pair of marbles always came together, and then remained together.

Close examination of this simple experiment revealed that upon modest agitation the small glass beads repeatedly bounced off of the marbles and off of each other, leading to a depletion of the bead density radially around each marble. When the marbles approached each other this excluded volume, now surrounding the closely spaced marbles, was less than the sum of the two excluded volumes when the marbles were further apart, consistent with the theory.

For a hypothetical idealized three-dimensional version of the glass bead and marble experiment (Figure 1) where the beads exhibit elastic collisions, the earth's gravitational field is absent, and there is no friction, it seems intuitively obvious that there should be an inverse square law for the force of attraction. Newton's inverse square law can be viewed as a consequence of having a constant number of gravitational field lines emanating radially from a massive spherical object and intersecting increasingly larger surface areas of imaginary spheres ($4\pi R^2$) drawn at increasing radius (R). The force thereby becomes weaker with increasing R by $1/R^2$ at any particular location, where a test mass is located [15]. A similar argument can be made for a spherical mass engulfed by agitated glass beads in three-dimensional space. The constant volume excluded to the small beads by a sphere causes a disturbance which radiates outward, and must influence increasingly larger surface areas, which increase by $R^2$ as R is increased, thereby diluting the influence of the beads with the inverse square of the radius at any particular location, as illustrated in Figure 2. Therefore, if empty space can be regarded as consisting of Planck volume-sized particles with quantum mechanical properties resembling those of the agitated glass beads, it would provide a physical mechanism for attraction by the "warping" of space by mass, consistent with general relativity. Of course, on the Planck scale macroscopic objects would appear to be mostly empty space, which would in general (except for black holes) be occupied by Planck volume-sized particles of space.

Finally, consider a spherical volume of empty space of radius R = 1 cm. Imagine that this spherical volume is packed at very high density with spherical unit Planck volumes each containing on the order of one bit of entropy.  The Planck volumes are undergoing considerable agitation due to quantum fluctuations, and move in a random fashion.  As mass is added to this macroscopic ($\approx$4 cm$^3$) volume, it displaces unit Planck volumes of space at the rate of approximately 1 bit of space entropy excluded per bit of mass entropy added.  Continuing this process until a black hole of R = 1 cm has formed excludes all of the Planck volumes, and the entropy change for this process would be zero.  One can imagine leaving a shell of only several Planck lengths thick encompassing the black hole.  This very thin empty shell volume would have an entropy content of approximately that of the surface area at R = 1 cm of the spherical black hole because R is such a large number ($10^{33}$) in Planck units.  If the entropy of the volume containing a black hole (with R from the black hole center being only a few Planck lengths larger than the black hole) is calculated, it would be approximately that of the surface area of the black hole, consistent with the Bekenstein-Hawking result. For lower-than-black-hole densities, this volume exclusion mechanism provides a simple explanation for the one-to-one correspondence between the information contained by a spherical volume of space and that passing through its surface area.  For each bit of information contained by a spherical volume, on the order of one bit of space entropy was displaced from the volume and passed through its surface area.  Thus, if the gravitational force arises from the exclusion of discrete Planck volume-size particles of entropy-containing space by matter, the holographic principle follows.  Moreover, the considerations presented here suggest that perhaps mass can be defined as the ability to displace space.

**ACKNOWLEDGEMENTS**

I would like to thank Ms. Yuanfan Ye for assistance with the experiments, and Dr. Minou Bina for helpful discussion.

**METHODS SUMMARY**

Marbles 2.5 cm in diameter were placed in a 20 cm diameter Pyrex glass dish containing approximately 37,500 (1 lb) 2 mm diameter glass beads and were shaken at 175 rpm in a standard rotary laboratory shaker. The marbles were submerged in the glass beads to a depth of approximately 5 mm. A plastic sheet containing a grid with 64 numbered one inch squares, placed on top of the dish, was used to record marble positions at the times indicated. Data similar to that of Figure 1 was obtained for many different marble starting positions in the dish for which the distance between the two marbles positions was from 10 to 13 cm (data not included). Shaking at speeds greater than approximately 250 rpm caused the marbles to rapidly approach each other, separate, and then approach each other again, repeatedly. Doubling the number of glass beads in the dish (and thereby increasing their weight to 2 lb) required more energy than could be required by the rotary shaker to sufficiently agitate the beads.

**FIGURE LEGENDS**

Figure 1. Demonstration of the attractive entropic force between two marbles partially submerged in small glass beads undergoing agitation by shaking. **a-h**, Frames of a movie taken at the times indicated. The plastic sheet containing a numbered grid covering the dish was used to record marble positions. The X's on the grid mark the starting positions of the darker (upper) and (mostly) lighter marbles, respectively. **i**, Distribution of the times that it took for the pair of marbles to come together in ten experiments. **j**, Illustration of the spiraling of one marble about the other that was usually observed. The numbers 0-3 adjacent to each marble (depicted here as black and white) denote three increasing times after the start of agitation. **k**, Average distances apart of the two marble positions at ten second intervals in ten experiments in which the marbles shaken either separately or together. The marbles were shaken under identical conditions from the same starting positions in the one-marble and the two-marble experiments. See **METHODS SUMMARY** for details of the experiments.

Figure 2. Cross sections of imaginary spheres centered on a massive spherical object M (gray filled circle) at radii of 1, 2, 3, and 4 units from the center of mass of M. The lighter gray color for each progressively larger circle reflects the decreasing disturbance of space (consisting of agitated Planck volumes) by the constant volume excluded by the mass on the surface area of each imaginary sphere, which increases with the square of the radius. The influence of the mass on the surrounding space eventually reduces to that of empty space.

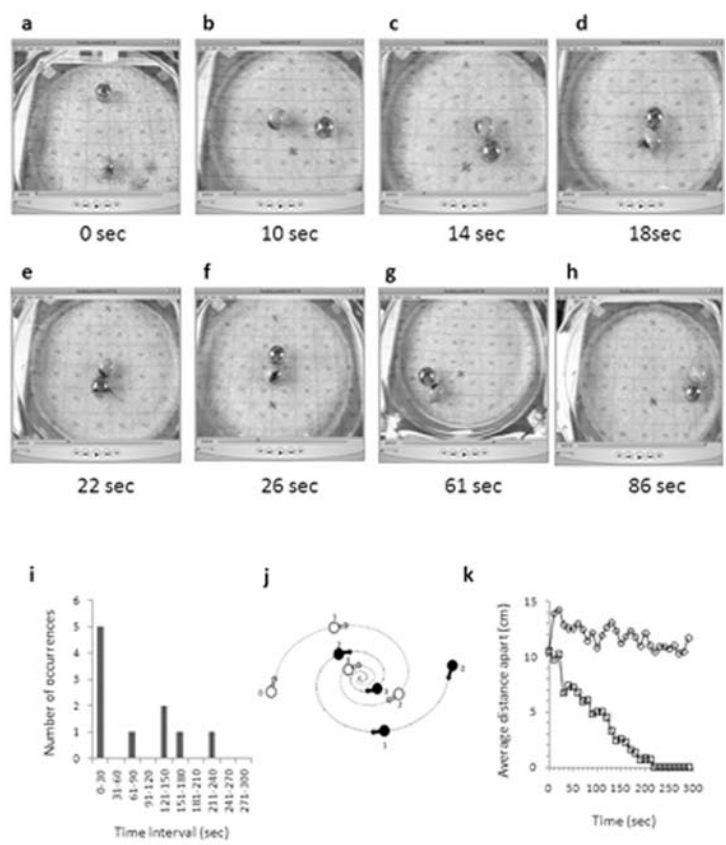

Fig. 1

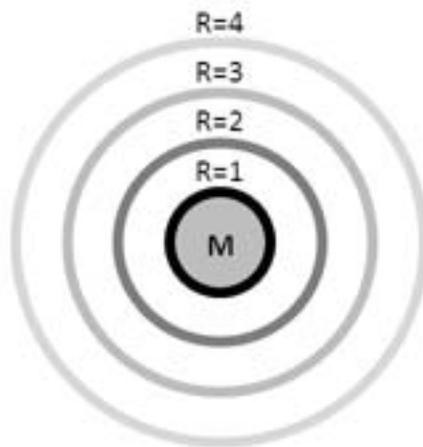

Fig. 2